\documentstyle[psfig]{mn} 
\title{Fundamental parameters of six neglected old open clusters}
\author[Carraro et al.]        
{Giovanni Carraro$^{1,2}$, \thanks{Andes Fellow, on leave from Dipartimento 
di Astronomia, Universit\`a di Padova,
Vicolo Osservatorio 2, I-35122, Padova, Italy}
Annapurni Subramaniam$^{3}$, and Kenneth A. Janes$^4$
\thanks{email: 
gcarraro@das.uchile.cl (GC), purni@iiap.res.in(AS), janes@bu.edu(KAJ)}\\ 
$^1$Departamento de Astronom\'ia, Universidad de Chile, 
Casilla 36-D, Santiago, Chile\\
$^2$Astronomy Department, Yale University, 
P.O. Box 208101, New Haven, CT 06520-8101 , USA\\
$^3$,Indian Institute of Astrophysics, II Block Koramangala, 
Bangalore 560034, India \\
$^4$Department of Astronomy, Boston University, 725 Commonwealth Avenue, Boston, MA 02215,
USA\\
} 
 
\date{\it Submitted: April 2006} 
\pubyear{2006} 
\begin{document} 
\maketitle 
\title{Basic parameters of old open clusters} 
            
\begin{abstract} 
In this paper we present the first $BVI$ CCD photometry of six overlooked old
open clusters (Berkeley~44, NGC~6827, Berkeley~52, Berkeley~56, Skiff~1 and Berkeley~5) 
and derive estimates of their fundamental parameters by using
isochrones from the Padova library (Girardi et al. 2000).
We found that all the clusters are older than the Hyades, with ages
ranging from 0.8 (NGC~6827 and Berkeley~5) to 4.0 (Berkeley~56) Gyr.
This latter is one of the old open clusters with the 
largest heliocentric distance.\\
In the field of Skiff~1 we recognize a faint blue Main Sequence identical to
the one found in the background of open clusters in the Second and Third
Galactic Quadrant, and routinely attributed to the Canis Major accretion event.
We use the synthetic Color Magnitude Diagram method and a Galactic model
to show that this population can be easily interpreted as Thick Disk
and Halo population toward Skiff~1. We finally revise the old open clusters
age distribution, showing that the previously suggested peak at 5 Gyr
looses importance as additional old clusters are discovered.
\end{abstract} 
 

\section{Introduction} 
The present day age distribution of open star clusters in the Galactic disk
is the result of the two competing processes: 
the star formation history of the Galactic disk and the dissolution
rate of star clusters (de la Fuente Marcos \& de la Fuente Marcos 2004).\\
The dissolution of star clusters is particularly important for the older
clusters, the typical open cluster lifetime being of the order of 200 Myr 
(Wielen 1971).
This way  to trace back the cluster formation history in the Galactic disk is 
a challenging task.
Recent compilations (Friel 1995, Ortolani et al. 2005) show that the age distribution of old open cluster has an e-folding shape with a possible
peak at 5 Gyrs. The reality of this peak is however quite difficult to assess,
and indeed a more recent analysis (Carraro et al. 2005, Fig.~10) shows that 
the inclusion
of a few overlooked clusters significantly weakens the reality of this peak,
and illustrates the importance of a careful hunting of old clusters before
drawing definitive conclusions.
The recent study of the old cluster Auner~1 (Carraro et al. 2006) with an age 
of 3.5 Gyr
again stresses the fact that we are still missing several old clusters.\\

Beginning with the paper of Phelps et al. (1994) several attempts have been 
made to enlarge the sample of studied old open clusters (Hasegawa et al. 2004, 
Carraro et al. 2005 and references therein).

In an attempt to further contribute to this interesting field,
in this paper we present the first photometric study
of 6 overlooked, old 
open clusters  
having $53^o\leq l \leq 130^o$ and 
$-5^o.2\leq b \leq +5^o.6$ (see Table~1) and provide homogeneous derivation
of basic parameters using the Padova (Girardi et al. 2000)
family of isochrones.\\
These clusters are NGC~6827, Berkeley 5, 52, 44, and 56 (Setteducati \& 
Weaver 1960)
and Skiff 1 (Luginbuhl \& Skiff 1990).

\noindent
The plan of the paper is as follows. Sect.~2 describes
the observation strategy and reduction technique.
Sect.~3 deals with star counts and radius determination.
The Color-Magnitude Diagrams (CMD) are described in Section~4, while
Section~5  illustrates the derivation of the clusters' fundamental
parameters. Section~6 concentrates on the star cluster Skiff~1.
Finally, Sect.~7 provides a detailed discussion
of the results.

\begin{table}
\caption{Basic parameters of the clusters under investigation.
Coordinates are for J2000.0 equinox and have been 
visually re-determined by us.}
\begin{tabular}{ccccc}
\hline
\hline
\multicolumn{1}{c}{Name} &
\multicolumn{1}{c}{$RA$}  &
\multicolumn{1}{c}{$DEC$}  &
\multicolumn{1}{c}{$l$} &
\multicolumn{1}{c}{$b$} \\
\hline
& {\rm $hh:mm:ss$} & {\rm $^{o}$~:~$^{\prime}$~:~$^{\prime\prime}$} & [deg] & [deg]\\
\hline
Berkeley 44    & 19:17:12 & +19:28:00 &  53.21 & +3.35\\
NGC 6827       & 19:48:54 & +21:12:00 &  58.25 & -2.35\\
Berkeley 52    & 20:14:18 & +28:58:00 &  67.89 & -3.13\\
Berkeley 56    & 21:17:42 & +41:54:00 &  86.04 & -5.17\\
Skiff 1        & 00:58:24 & +68:28:00 & 123.57 & +5.60\\
Berkeley 5     & 01:47:48 & +62:56:00 & 129.29 & +0.76\\
\hline\hline
\end{tabular}
\end{table}

\begin{figure} 
\centerline{\psfig{file=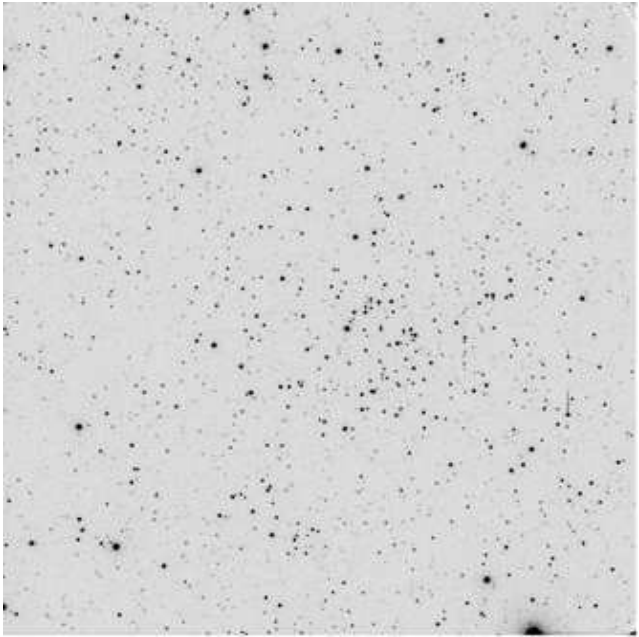,width=\columnwidth}} 
\caption{V  60 secs image centered on Berkeley 44.
North is up, east on the left. The field is 10 arcmin
on a side.}
\end{figure} 

\begin{figure} 
\centerline{\psfig{file=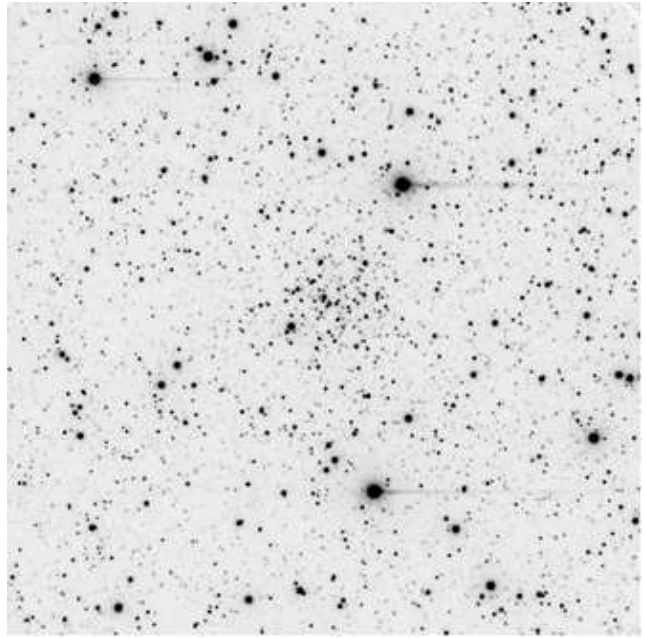,width=\columnwidth}} 
\caption{V  60 secs image centered on NGC 6827.
North is up, east on the left.}
\end{figure}

\begin{figure} 
\centerline{\psfig{file=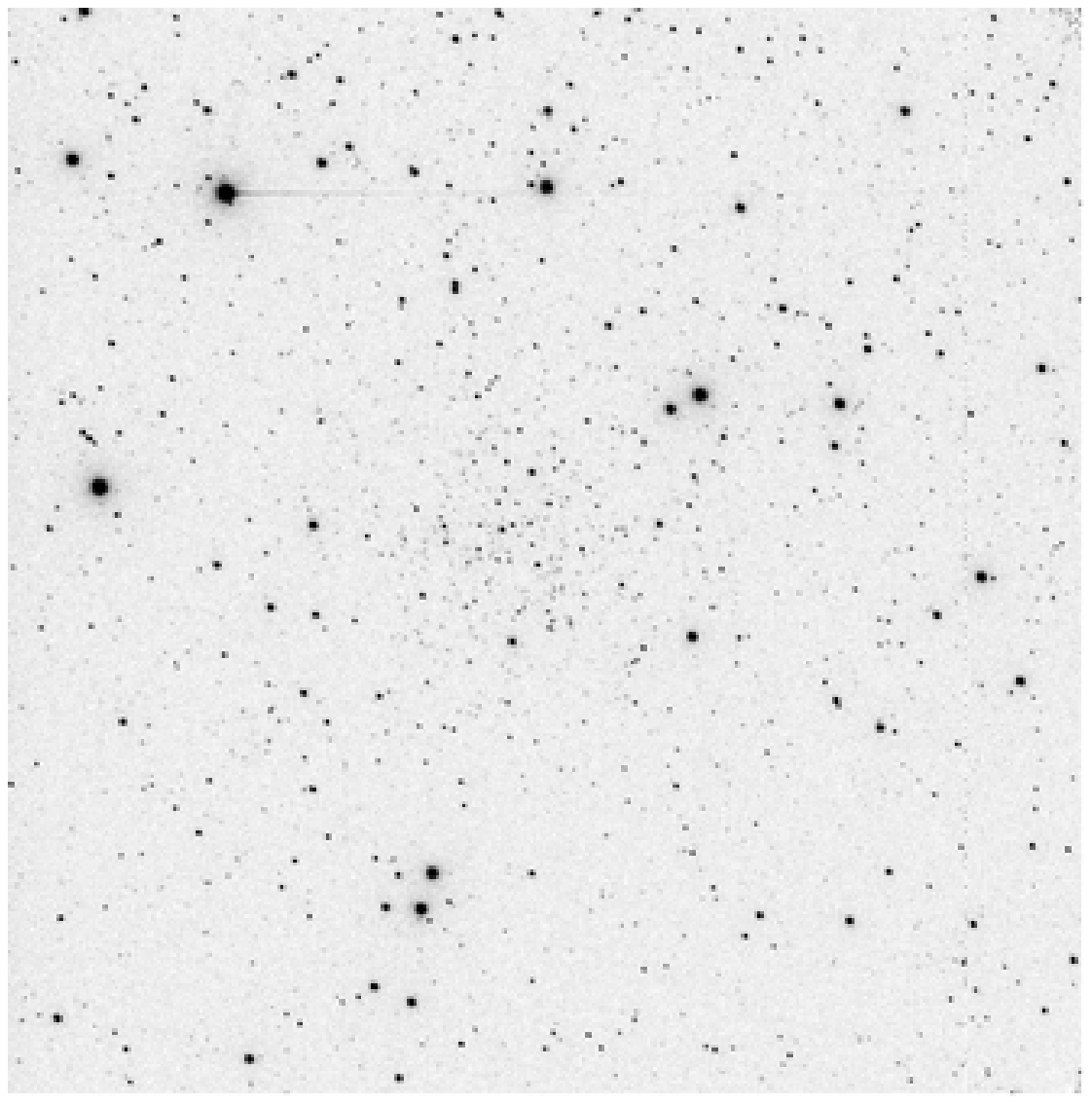,width=\columnwidth} }
\caption{V  60 secs image centered on Berkeley 52.
North is up, east on the left.}
\end{figure} 

\begin{figure} 
\centerline{\psfig{file=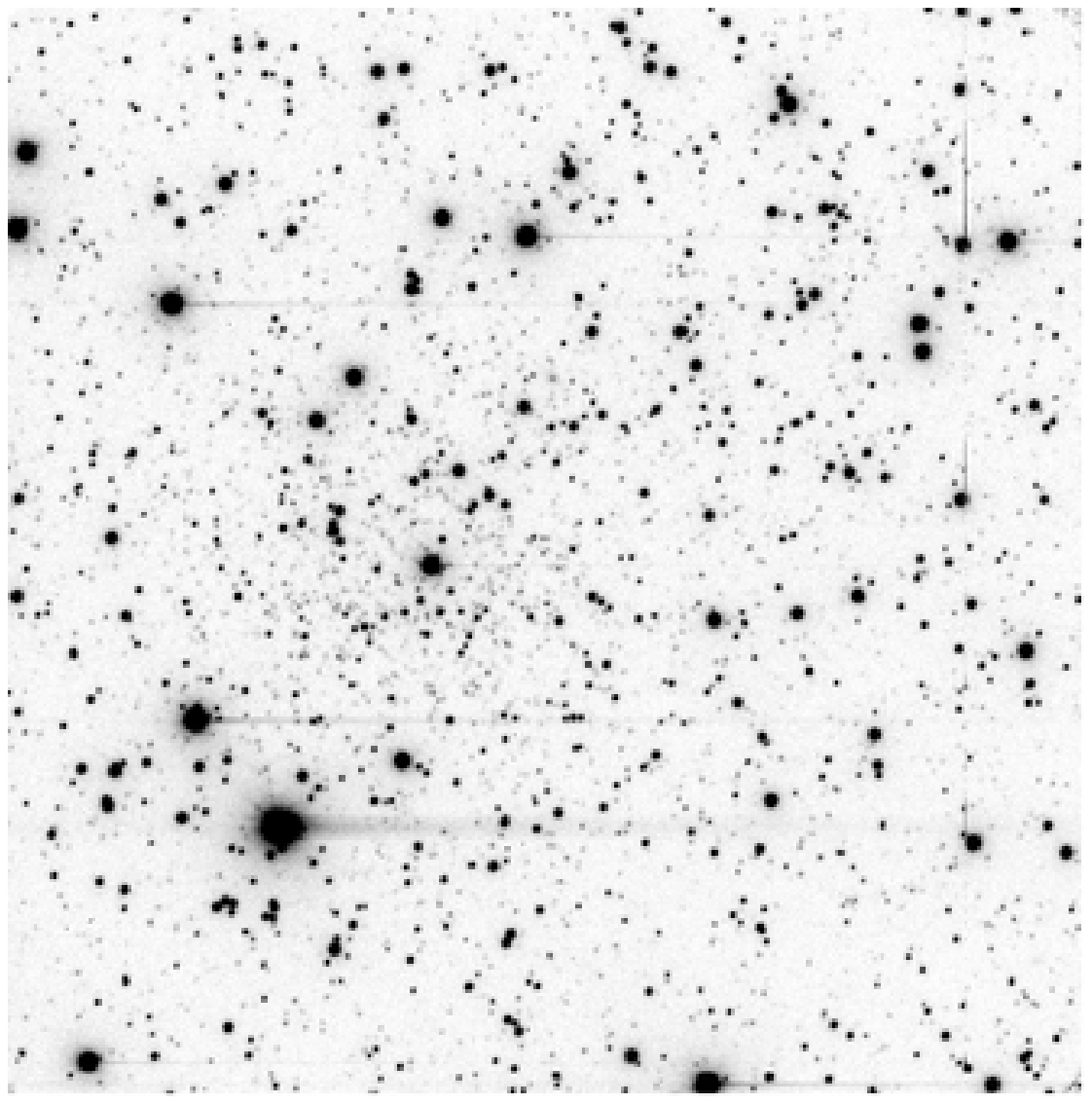,width=\columnwidth}} 
\caption{V  60 secs image centered on Berkeley 56.
North is up, east on the left.}
\end{figure}

\begin{figure} 
\centerline{\psfig{file=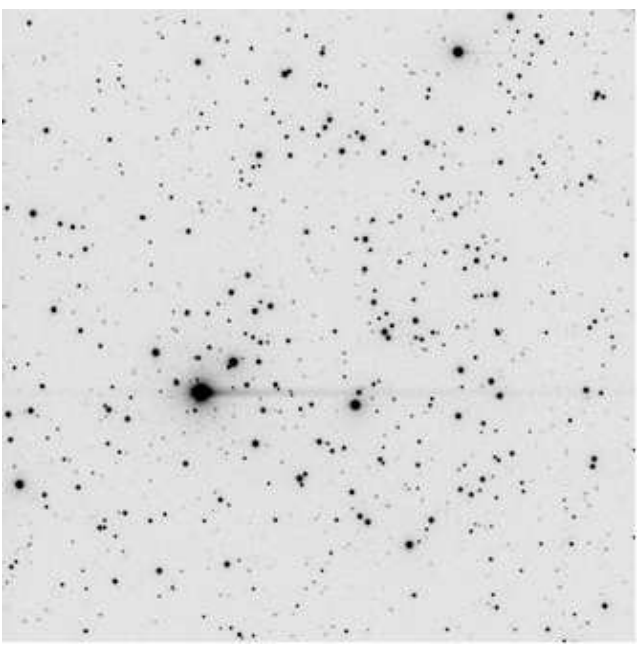,width=\columnwidth} }
\caption{V  60 secs image centered on Skiff 1.
North is up, east on the left.}
\end{figure} 

\begin{figure} 
\centerline{\psfig{file=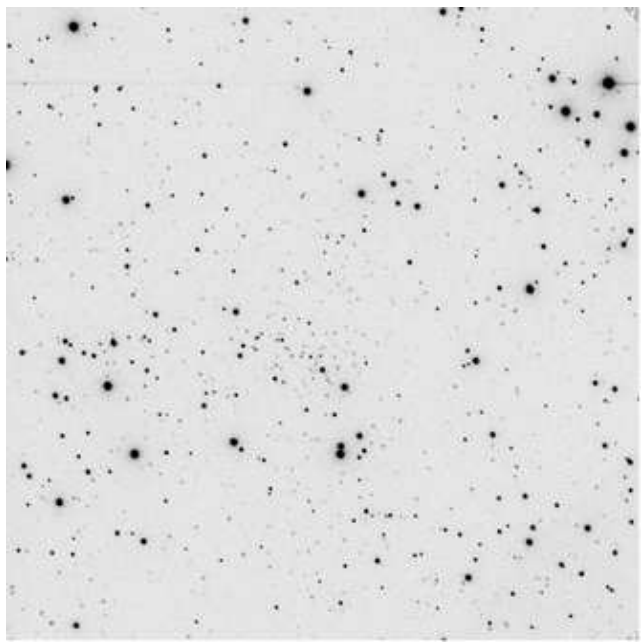,width=\columnwidth} }
\caption{V  60 secs image centered on Berkeley 5.
North is up, east on the left.}
\end{figure}

\section{Observations and Data Reduction} 

The observations were done using the 2-m Himalayan Chandra Telescope (HCT),
located at Hanle, IAO and operated by Indian Institute of Astrophysics.
Details of the telescope and the instrument are available at the
Institute's homepage (http://www.iia.res.in/).
The CCD used for imaging is a 2 K $\times$ 4 K CCD,
where the central 2 K $\times$ 2 K pixels were used for imaging. The pixel 
size is 15 $\mu$
with an image scale of 0.297 arcsec/pixel and the average seeing
was 1.3 and 1.4 arsec on August 9 and 30, respectively. 
The total area observed is
approximately 10 $\times$ 10 arcmin$^2$. Images of the clusters are presented 
in Figures 1 - 6.  \\

\noindent
The data have been reduced with the
IRAF\footnote{IRAF is distributed by NOAO, which are operated by AURA under
cooperative agreement with the NSF.}
packages CCDRED, DAOPHOT, ALLSTAR and PHOTCAL using the point spread function 
(PSF) method (Stetson 1987).\\

The nights were photometric and Landolt (1992)
standard field SA110 was observed for calibration at different air-masses
during the night to put the photometry into the standard system.

\begin{table}
\centering
\caption{Log of photometric observations on August 9 and 30, 2005.}
\begin{tabular}{lrrr}
\hline
Cluster& Date & Filter & Exp time (sec) \\
\hline
Be 44 & 09 August 2005 & V & 60, 180, 2X300\\
      &                & B & 2X120, 2X600\\
      &                & I & 2, 5, 10, 2X30, 2X60\\
Be 44 (Field) &        & V & 60, 2X300\\
              &        & B & 2X600\\
              &        & I & 2, 2X60\\
NGC 6827 & 30 August 2005 & V &60, 180, 2X300\\
         &                & B & 120, 300, 2X600\\
         &                & I & 10, 30, 60, 2X120\\
Be 52 & 09 August 2005 & V & 60, 180, 2X420\\
      &                & B & 120, 600, 900\\
      &                & I & 10, 30, 120, 300 \\
Be 56 & 30 August 2005 & V & 30, 60, 2X180\\
      &                & B & 180, 300, 600 \\
      &                & I & 10, 30, 60, 2X120\\
Skiff 1 & 30 August 2005 & V & 20, 60, 2X180\\
        &                & B & 30, 120, 300, 600\\
        &                & I & 10, 30, 60, 120\\
Skiff 1 (east) &         & V &30, 180 \\
               &         & B & 60, 300\\
               &         & I & 30, 120\\
Be 5 & 30 August 2005 & V & 60, 3X180\\
     &                & B & 60, 300, 600 \\
     &                & I & 10, 30, 180, 300 \\
\hline
\end{tabular}
\end{table}
\begin{figure} 
\centerline{\psfig{file=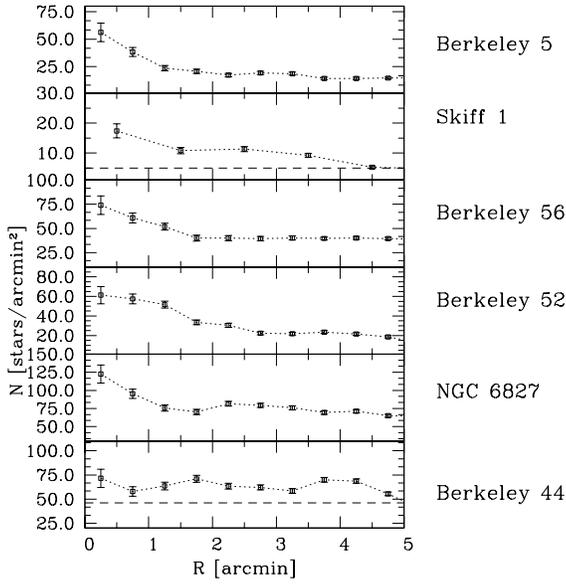,width=\columnwidth}} 
\caption{Star counts in the V passband for the clusters under investigation.
{\bf The dashed lines  in the Berkeley~44 and Skiff~1 panels
indicate the level of the background as derived from
the accompanying control field.}}
\end{figure}

\begin{figure} 
\centerline{\psfig{file=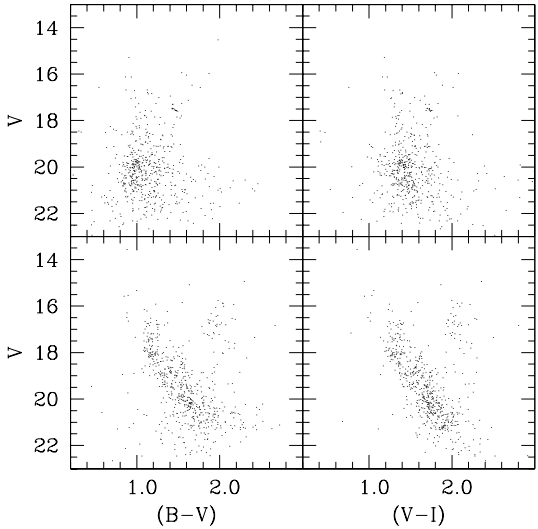,width=\columnwidth}} 
\caption{Color Magnitude Diagrams of NGC 6827 {\bf(lower panels)}
and Berkeley 56 {\bf (upper panels)}.
Only the stars inside the cluster radius are shown.}
\end{figure} 

\begin{figure} 
\centerline{\psfig{file=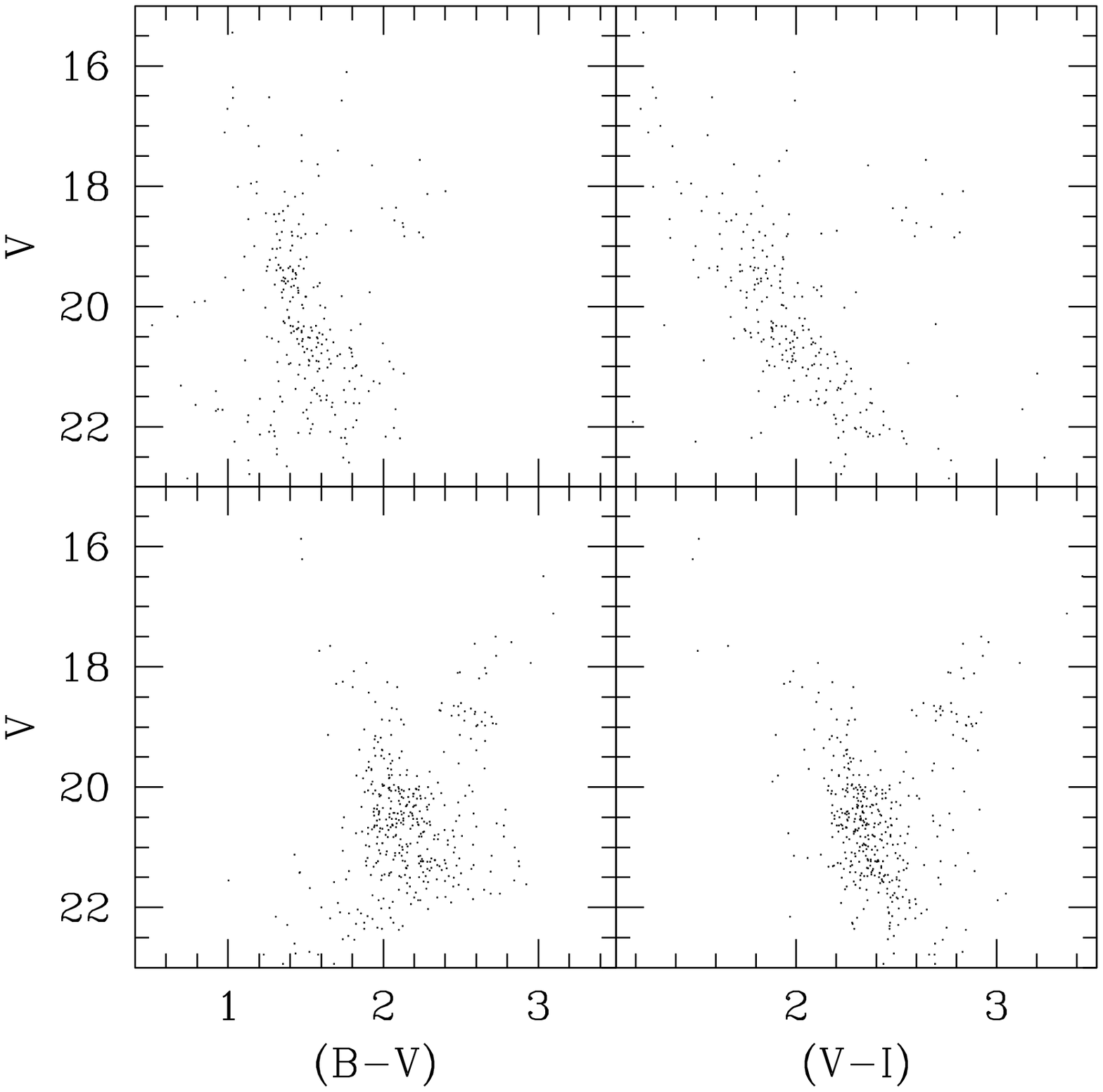,width=\columnwidth}} 
\caption{Color Magnitude Diagrams of Berkeley~52 {\bf(lower panels)}  
and Berkeley 5 {\bf (upper panels)}.
Only the stars inside the cluster radius are shown.}
\end{figure}

Together with the clusters, 
we observed two control fields, one
east of Skiff~1 at 01:06:24, +68:29:00 (J2000.0), and the other 
north of Berkeley~44
at 19:17:12, +19:38:00 (J2000.0), to deal with field star
contamination. 
In fact these are the only two clusters which seem to extend
beyond the field covered by the CCD.\\

\noindent
The calibration equations are of the form:\\

\noindent
$ b = B + b_1 + b_2 \times X + b_3~(B-V)$ \\
$ v = V + v_1 + v_2 \times X + v_3~(B-V)$ \\
$ i = I + i_1 + i_2 \times X + i_3~(V-I)$ ,\\

\begin{table}
\tabcolsep 0.2truecm
\caption {Coefficients of the calibration equations: August 9, 2005}
\begin{tabular}{ccc}
\hline

$b_1 = 0.803 \pm 0.007$ & $b_2 =  0.25 \pm 0.02$ & $b_3 = -0.043 \pm 0.006$ \\
$v_1 = 0.495 \pm 0.006$ & $v_2 =  0.16 \pm 0.02$ & $v_3 =  0.063 \pm 0.004$ \\
$i_1 = 0.826 \pm 0.012$ & $i_2 =  0.08 \pm 0.02$ & $i_3 =  0.044 \pm 0.009$ \\

\hline
\end{tabular}
\end{table}

\begin{table}
\tabcolsep 0.2truecm
\caption {Coefficients of the calibration equations: August 30, 2005}
\begin{tabular}{ccc}
\hline

$b_1 = 0.818 \pm 0.008$ & $b_2 =  0.26 \pm 0.02$ & $b_3 =  -0.043 \pm 0.008$ \\
$v_1 = 0.513 \pm 0.005$ & $v_2 =  0.14 \pm 0.02$ & $v_3 =  0.063 \pm 0.005$ \\
$i_1 = 0.824 \pm 0.009$ & $i_2 =  0.08 \pm 0.02$ & $i_3 =  0.048 \pm 0.009$ \\

\hline
\end{tabular}
\end{table}

\noindent
where $BVI$ are standard magnitudes, $bvi$ are the instrumental ones and  
$X$ is
the airmass; all the coefficient values are reported in Tables~3 and 4.
The standard stars in these fields provide a very good color coverage
being 0.1 $\leq (B-V) \leq $ 2.2 and 0.4 $\leq (V-I) \leq $ 2.6\\

\noindent
Aperture correction was then derived from a sample of bright stars
and applied to the photometry.  We used aperture of 14 pixels for the standards
stars and of 7-9 pixels for the science frames, depending on the frame.
The average aperture correction amounted at
0.27, 0.29 and 0.20 mag in B,V and I, respectively for the August 9
night, and 0.25, 0.25 and 0.21 for the August 30 night.\\

Finally, the completeness corrections were determined by artificial-star 
experiments
on our data. Basically,
we created several artificial images by adding to the original images 
artificial
stars. About a total of 4000 stars were added to the original images.
In order to avoid the creation of overcrowding, in each experiment we added
at random positions only 15$\%$ of the original number of stars. The artificial
stars had the same color and luminosity distribution of the original sample.
This way we found that the completeness level keeps above 50$\%$ down to 
V = 20.5.

\noindent
The limiting magnitudes are B = 22.0, V = 22.5
and I =21.5.\\
The final photometric catalogs for
(coordinates,
B, V and I magnitudes and errors)
consist of 11000, 10525, 12730, 2250, 2973, 7117, and 6486 stars
for NGC~6827, NGC~6846, Berkeley~44, Berkeley~5, Berkeley~52, Berkeley~56
and Skiff~1, respectively, and are made
available in electronic form at the
WEBDA\footnote{http://www.univie.ac.at/webda/navigation.html} site
maintained by E. Paunzen.\\

\section{Star counts and cluster sizes} 
As a first step in the analysis of the clusters, we performed
star counts to obtain an estimate of the cluster radius. This is
an important step in order to pick up the most probable cluster
members and minimize field star contamination.
By inspecting clusters charts we identified the cluster center,
and performed star counts in circular annuli 0.5 arcmin wide around
the cluster center. In order to increase the contrast, we
consider in each cluster only the stars fainter than the clump.\\ 
The results are shown in Fig. 7. Here the
error bars are the Poisson error of the star counts in each annulus.\\
{\bf In the case of Berkeley~44 and Skiff~1 we estimate the level
of the background from the accompanying offset field, and draw
it with a dashed line in Fig.~7.}
\noindent
By inspecting Fig.~7 the following considerations can be done:

\begin{itemize}
\item NGC~6827, NGC~6846, Berkeley~52, Berkeley~56 and Berkeley~5 are compact 
clusters
with radii between 1 and 2 arcmin;
\item Berkeley~44 does not show a well-defined outer radius, since star counts 
fall smoothly along the entire area covered by this study. For this cluster
we have observed an offset field (see Sect.~2) which we are going to analyze 
in the following Section.
\item Although there is a visible overdensity of stars in the Skiff~1 region
the overdensity  suggests
the form of a ring structure 2 arcmin from the cluster nominal center. As
shown in the following discussion, Skiff~1 is both a sparse cluster
and rather
more nearby than the 
other clusters. For this reason we have used a different annulus size (1.0 arcmin)
to derive the profile.   
\end{itemize}

\noindent
Estimates of the cluster sizes, taken from Figure 7, are presented in Table 6;
they are in good agreement with the Dias et al. (2002)
compilation, which is based on visual inspection.

\begin{figure*} 
\centerline{\psfig{file=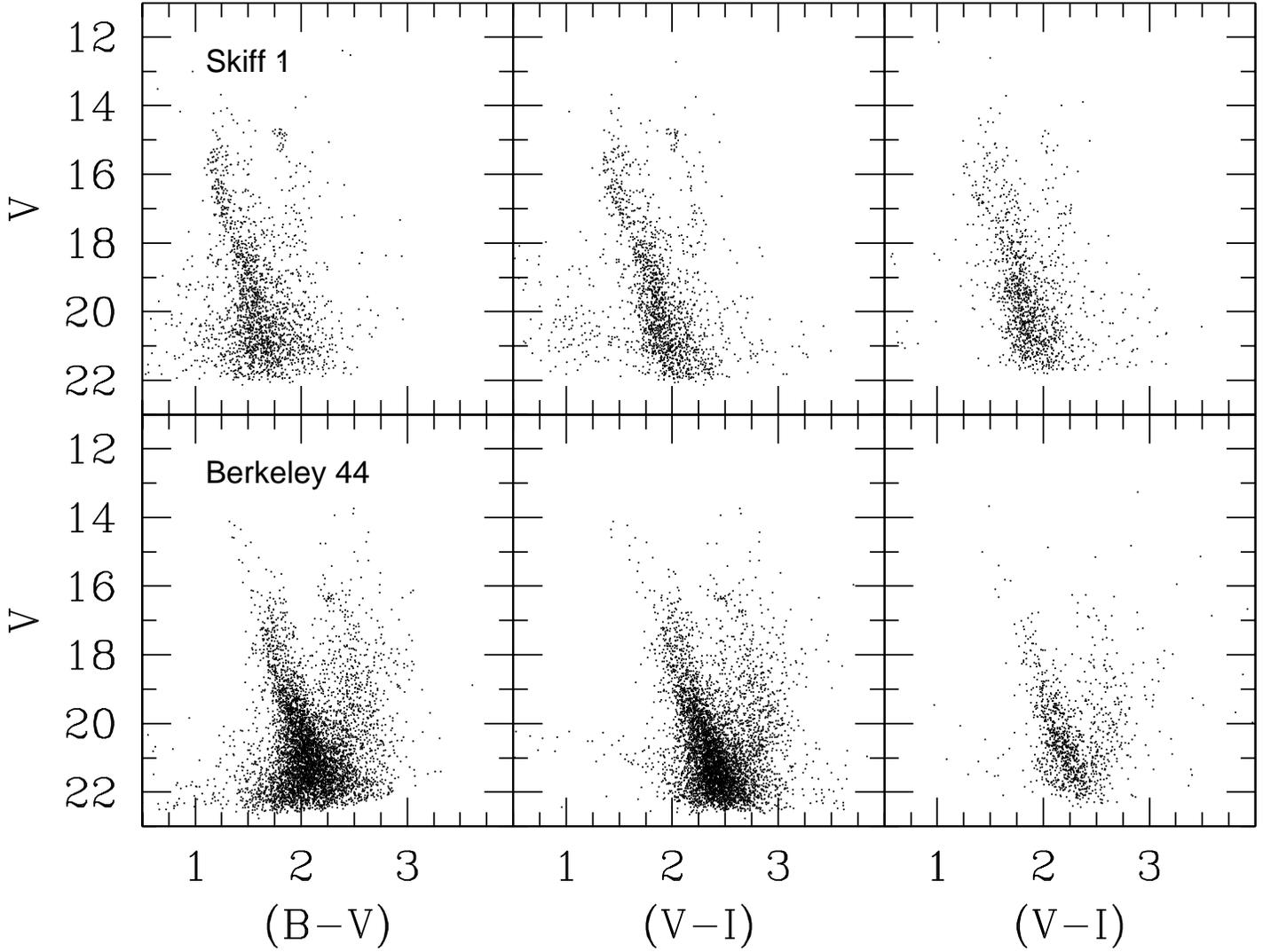}} 
\caption{Color Magnitude Diagrams of Skiff 1 and Berkeley 44.
The right panels are the comparison field. Note the anomalous
blue MS with a TO at V = 19.5 in the CMD of Skiff 1 and its control field. }
\end{figure*} 

\begin{table*}
\caption{Preliminary age estimates based on the $\Delta~V$ method}
\fontsize{8} {10pt}\selectfont
\begin{tabular}{ccccccccc}
\hline
\multicolumn{1}{c} {$Name$} & 
\multicolumn{1}{c} {$V_TO$} &
\multicolumn{1}{c} {$(B-V)_{TO}$}  &
\multicolumn{1}{c} {$(V-I)_{TO}$} &
\multicolumn{1}{c} {$V_{clump}$} &
\multicolumn{1}{c} {$(B-V)_{clump}$} &
\multicolumn{1}{c} {$(V-I)_{clump}$} &
\multicolumn{1}{c} {$\Delta V$} &
\multicolumn{1}{c} {$Age$} \\
\hline
& mag & mag & mag& mag & mag & mag & mag & Gyr \\
\hline
Berkeley~44  & 17.50$\pm$0.05 & 1.75$\pm$0.10 & 2.00$\pm$0.10 & 16.50$\pm$0.11 & 2.25$\pm$0.12 & 2.50$\pm$0.14  & 1.00$\pm$0.12  & 1.1$\pm$0.25\\
NGC~6827     & 17.50$\pm$0.05 & 1.20$\pm$0.10 & 1.40$\pm$0.10 & 16.75$\pm$0.25 & 2.00$\pm$0.29 & 2.15$\pm$0.32  & 0.75$\pm$0.25  & 0.8$\pm$0.20\\
Berkeley~52  & 20.50$\pm$0.05 & 2.00$\pm$0.10 & 2.20$\pm$0.10 & 19.00$\pm$0.09 & 2.50$\pm$0.23 & 2.80$\pm$0.25  & 1.50$\pm$0.10  & 1.8$\pm$0.30\\
Berkeley~56  & 20.50$\pm$0.05 & 0.80$\pm$0.10 & 1.10$\pm$0.10 & 17.70$\pm$0.08 & 1.50$\pm$0.16 & 1.75$\pm$0.18  & 2.30$\pm$0.09  & 4.0$\pm$0.50\\
Skiff~1      & 15.50$\pm$0.05 & 1.10$\pm$0.10 & 1.30$\pm$0.10 & 14.70$\pm$0.11 & 1.80$\pm$0.13 & 2.00$\pm$0.13  & 0.80$\pm$0.12  & 0.9$\pm$0.20\\
Berkeley~5   & 19.50$\pm$0.05 & 1.30$\pm$0.10 & 1.50$\pm$0.10 & 18.60$\pm$0.17 & 2.10$\pm$0.21 & 2.40$\pm$0.23  & 0.90$\pm$0.18  & 1.0$\pm$0.20\\
\hline
\end{tabular}
\end{table*}

\section{Color Magnitude Diagrams: Are these real clusters?}
By using the results of the previous section we generate
the CMDs of the clusters considering only the stars
within the assumed cluster radius (Table 6). The results 
are shown in
Figs. 8 to 10. All of the clusters are located in crowded galactic
plane fields and the CMDs are heavily contaminated with the projected 
background main sequence population of the galaxy.  In spite of this
contamination, an apparent red giant clump is
noticeable on all of the diagrams.  We use this as our first evidence for the 
existence of physical clusters.

\noindent
We can improve the contrast between the clusters and the background field by 
employing a statistical method to clean the CMDs.
For each cluster, we selected a field region far from the cluster region. 
This selection
was done in in the same CCD field for all the clusters except
Berkeley~44 and Skiff~1, for which we have at disposal an offset field.
The cluster and field regions have the same area.\\

\noindent
To perform the statistical subtraction, we employed
the technique described in Vallenari et al. (1992)
and Gallart et al. (2003).\\
Briefly, for any star in the field, we look for the closest
(in color and magnitude) star in the cluster, and remove
this star from the cluster CMD. This procedure takes into account
the photometric completeness (see Section~2.)\\
\noindent
The results are shown in the series of Figs. 11 to 16.
In these figures, the left panel shows the CMD for stars inside the 
selected radius, whereas the mid-left panel
shows the offset equal area field.
The {\it cleaned} CMD is then shown in the mid-right panel.  In each case, the
cleaning process leaves an apparent cluster CMD; we assume in the remainder
that all of these are physical systems. 

Finally the isochrone fitting is presented in the right panel (see the next 
section).\\

\noindent
To get a first estimate of cluster age, we now employ the 
$\Delta V$ (magnitude difference between
the TO and the RGB clump) vs age calibration by Carraro $\&$ Chiosi (1994). 
This method is independent
of distance and reddening, and depends only on metallicity.
The results are summarized in Table~5  together with
their uncentainties.
The magnitude and colors of the TO have been estimated by eye, whilst
the magnitude and colors of the clump are the mean magnitude and colors of the
stars in the clump area in the CMD.
Basing on this method all the clusters are of Hyades age or older,
with Berkeley~56 being the oldest of the sample.

\noindent
An inspection of each CMD allows us to
derive the following considerations:\\

\noindent
{\bf NGC~6827}. The cluster looks like an intermediate-age
one, with a prominent clump of stars at V $\approx$ 16.5 and (B-V)
$\approx$2.7, (V-I)  $\approx$ 2.1. The Turn Off point (TO)
is located at V  $\approx$ 17.75, (B-V)  $\approx$ 1.0.
{\bf The MS looks truncated at V$\sim$ 19.5 as a result
of the cleaning procedure}\\

\noindent
{\bf Berkeley~52}. This is a faint and heavily reddened cluster.
The presence of a clear clump witnesses that the cluster is relatively old.\\
 
\noindent
{\bf Berkeley~5} This cluster is poorly populated; the clump, if real,
is very sparse, which can be a signature of significant differential
reddening. The TO however is readily detectable, which ensures
the reality of this cluster.\\

\noindent
{\bf Berkeley~56}  It looks a promising old cluster, with a tight clump at 
V $\approx$ 17.5. The TO area is at the limit of the photometry, although
the TO can easily be identified at V $\approx$ 20.\\

\noindent
{\bf Skiff~1} This is a very interesting object. There is clear clump at 
V $\approx$ 15,
which is not visible in the control field and 
ensures this is a real intermediate-age/old
cluster. We have to note here (both in the cluster and the offset field) the presence of a faint blue
population with a TO at V = 19.5. 
This is similar to the one detected in the third Galactic Quadrant
(Bellazzini et al. 2004)
and in the second Galactic Quadrant
(Bragaglia et al. 2006) and routinely attributed to the Canis Major Galaxy 
(Bellazzini et al. 2004).   
This is quite remarkable since this presumed dwarf Galaxy (or its tidal tail) 
is not expected to extend to this Galactic location (Martin et al. 2004, 
Fig.~4).\\

\noindent
{\bf Berkeley~44} The contamination of field stars is severe in this case.
However the upper part of the MS and the evolved stars region
are significantly more populated than in the control field.
We suggest that this cluster is real.\\

\section{Estimates of Fundamental parameters}
In order to derive more reliably 
the cluster fundamental parameters, namely reddening, age
and distance, we employ the following technique, using the {\it cleaned} CMDs
in figures 11 to 16.

\noindent
Possible isochrone solutions are obtained by exploring a large number
of isochrones; for clarity only the isochrone with the best visual match to the
observed one is presented.
To achieve the best match we paid attention to the slope of the MS,
the position and shape of the TO region and the magnitude and color of the
RGB clump. These constraints are function of age, metallicity,
distance and reddening, and must be reproduced at the same time.
Lacking a spectroscpopic estimate of the metallicity, we employ
the solar metallicity set.

\begin{figure*} 
\centerline{\psfig{file=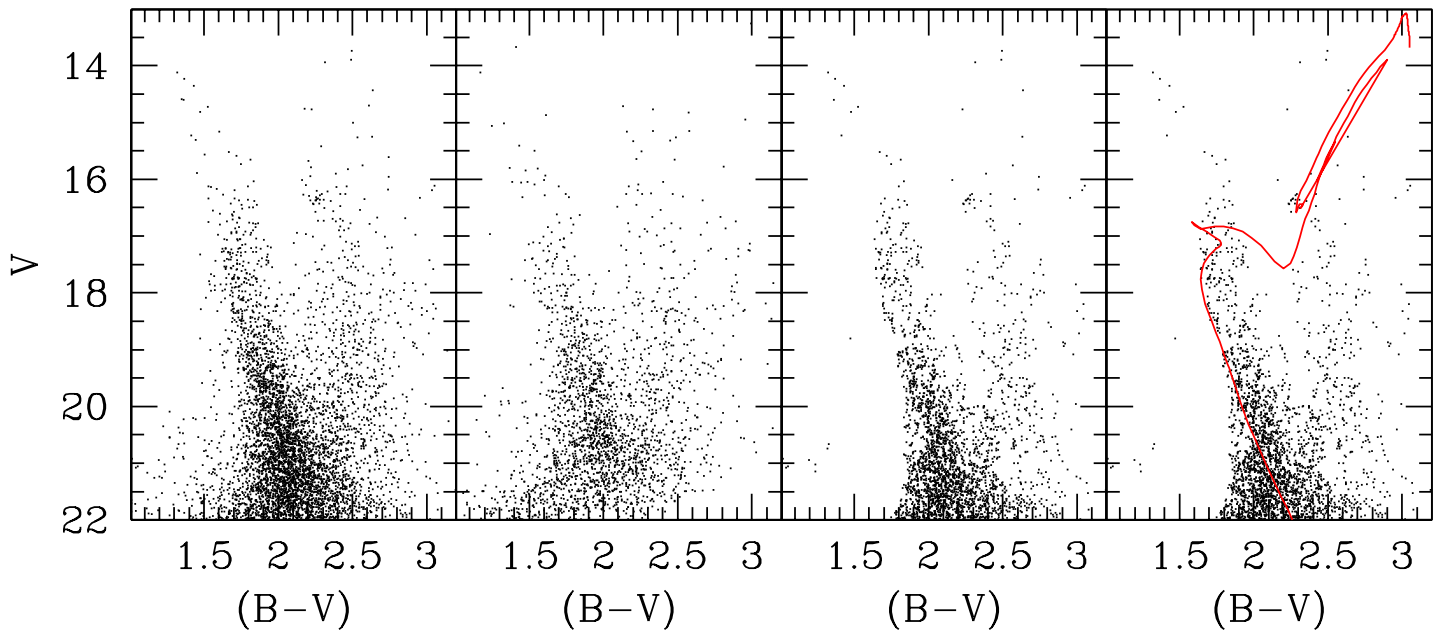} }
\caption{{\bf Left panel:} CMD of Berkeley~44;{\bf Mid-left panel:}
CMD of the control field; {\bf Mid-right panel:} the clean CMD;
{\bf Right  panel:}Isochrone solution for Berkeley 44: the 1.3 Gyr
isochrone is shitfed by E(B-V) = 1.40 en V-M$_V$ = 15.60.}
\end{figure*} 

\begin{figure*} 
\centerline{\psfig{file=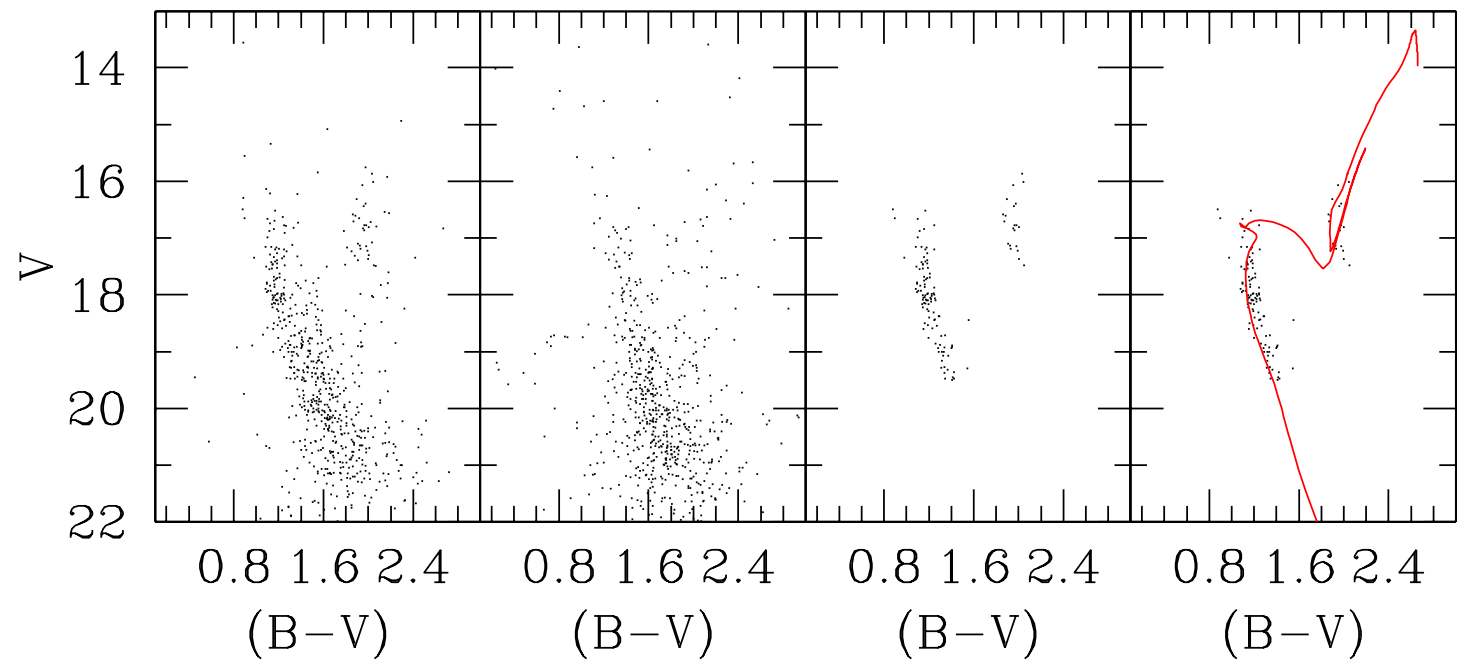} }
\caption{{\bf Left panel:} CMD of NGC~6827;{\bf Mid-left panel:}
CMD of the control field; {\bf Mid-right panel:} the clean CMD;
{\bf Right  panel:}Isochrone solution for NGC~6827: the 0.8 Gyr
isochrone is shitfed by E(B-V) = 1.05 en V-M$_V$ = 16.30}
\end{figure*} 

\begin{figure*} 
\centerline{\psfig{file=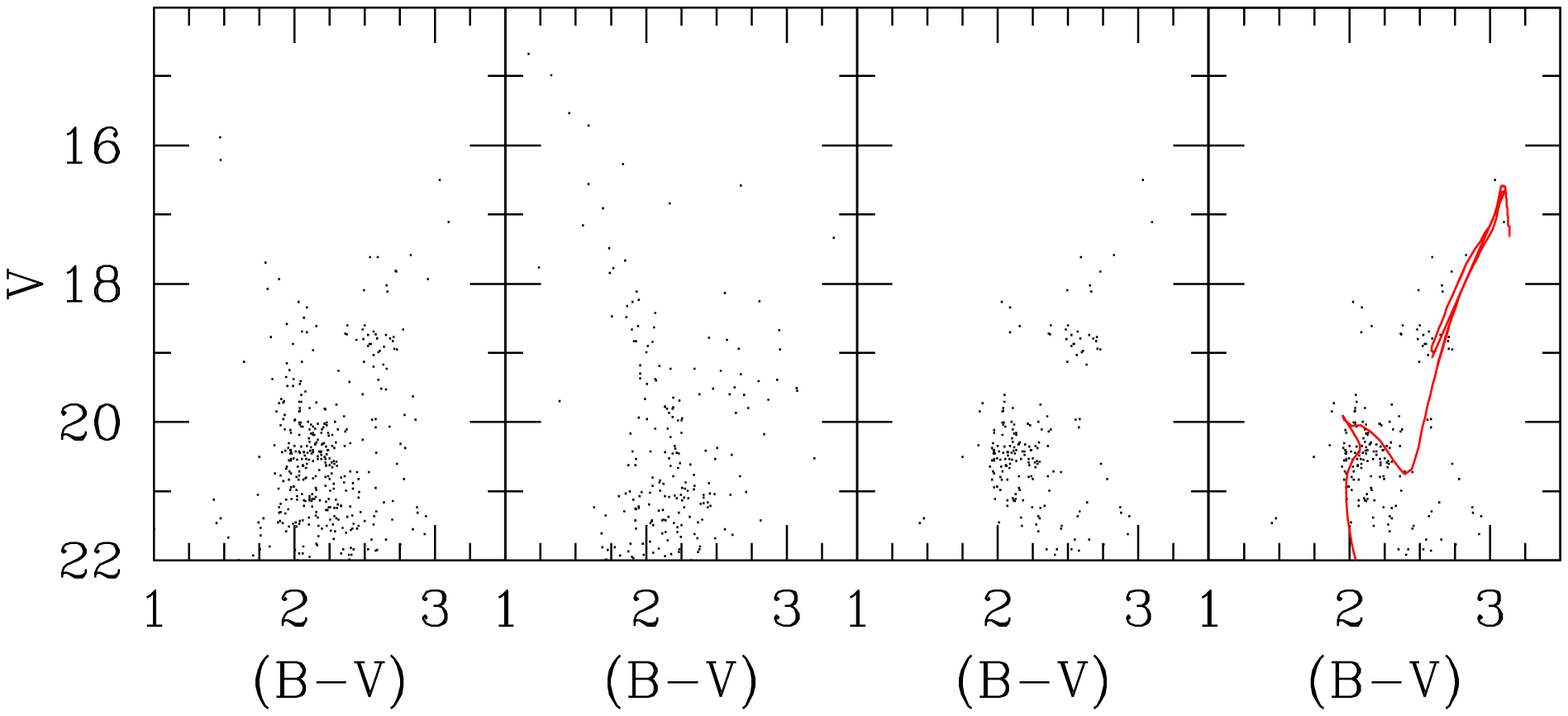} }
\caption{{\bf Left panel:} CMD of Berkeley~52;{\bf Mid-left panel:}
CMD of the control field; {\bf Mid-right panel:} the clean CMD;
{\bf Right  panel:}Isochrone solution for Berkeley~52: the 2.0 Gyr
isochrone is shitfed by E(B-V) = 1.50 en V-M$_V$ = 18.10}
\end{figure*} 

The uncertainty in each parameter simply mirrors the degree
of freedom we have in displacing an isochrone still achieving
an acceptable fit.

\begin{figure*} 
\centerline{\psfig{file=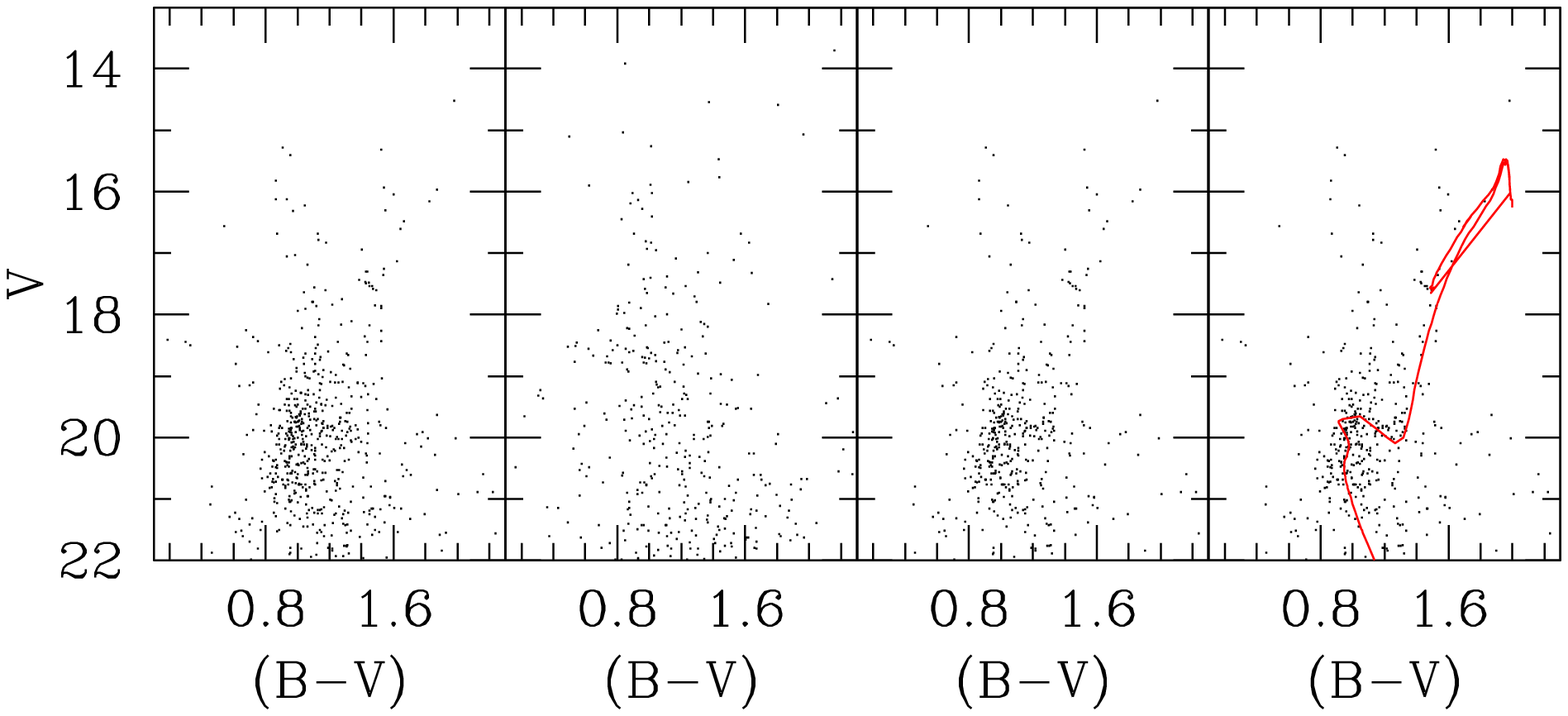} }
\caption{{\bf Left panel:} CMD of Berkeley~56;{\bf Mid-left panel:}
CMD of the control field; {\bf Mid-right panel:} the clean CMD;
{\bf Right  panel:}Isochrone solution for Berkeley~56: the 4.0 Gyr
isochrone is shitfed by E(B-V) = 0.40 en V-M$_V$ = 16.60}
\end{figure*} 

The results of the isochrone fitting method
are summarized in Table~6, where for each cluster 
radius, reddening, distance modulus, heliocentric
distance, Galactic Cartesian coordinates, Galactocentric
distance, and age are reported.\\
To derive the cluster heliocentric distance we corrected
the apparent distance modulus (V-M$_V$) by adopting the standard
ratio of selective to total absorption $R_V = \frac{A_V}{E(B-V)}= 3.1$.\\

\noindent
A few comments are in order:

\begin{itemize}
\item All the clusters are substantially reddened, with E(B-V) ranging from
0.40 to 1.50;
\item  all the cluster are older than the Hyades, and therefore they
constitute a significant contribution to the old open cluster population
in the Galactic disk;
\item  Two of them lie inside the solar circle, which is a remarkable result,
since star clusters are not expected to survive so long in the dense 
environment
typical of the inner part of the Galactic disk;
\item they span about 7 kpc in Galactocentric distance, but they do not seem
to follow the radial abundance gradient (Carraro et al. 1998); this however
has to be considered a preliminary results, due to the really crude estimate
of the metallicity we can infer from isochrone fitting.
\item the oldest cluster of the sample is Berkeley~56, which also lies high 
onto
the Galactic plane, and it is one of the most distant cluster from the Sun  
(Friel 1995)
\item Berkeley~56 with an age of 4 Gyr falls in an age bin where a minimum
in the star cluster age distribution was suggested to exist; however the 
discovery
of several new clusters in this age bin (Carraro et al. 2005) significantly
reduces the reality of this minimum, and suggests that the age distribution
of old open clusters is simply an e-folding relation.
\item the clean  CMDs of Berkeley 44, 52 and 56 show the presence of a bunch
of stars above the TO. These stars can be field stars that the cleaning procedure
was not able to remove, or they can be blue stragglers and binary stars, quite
common in clusters of this age.
\end{itemize}

\begin{figure*} 
\centerline{\psfig{file=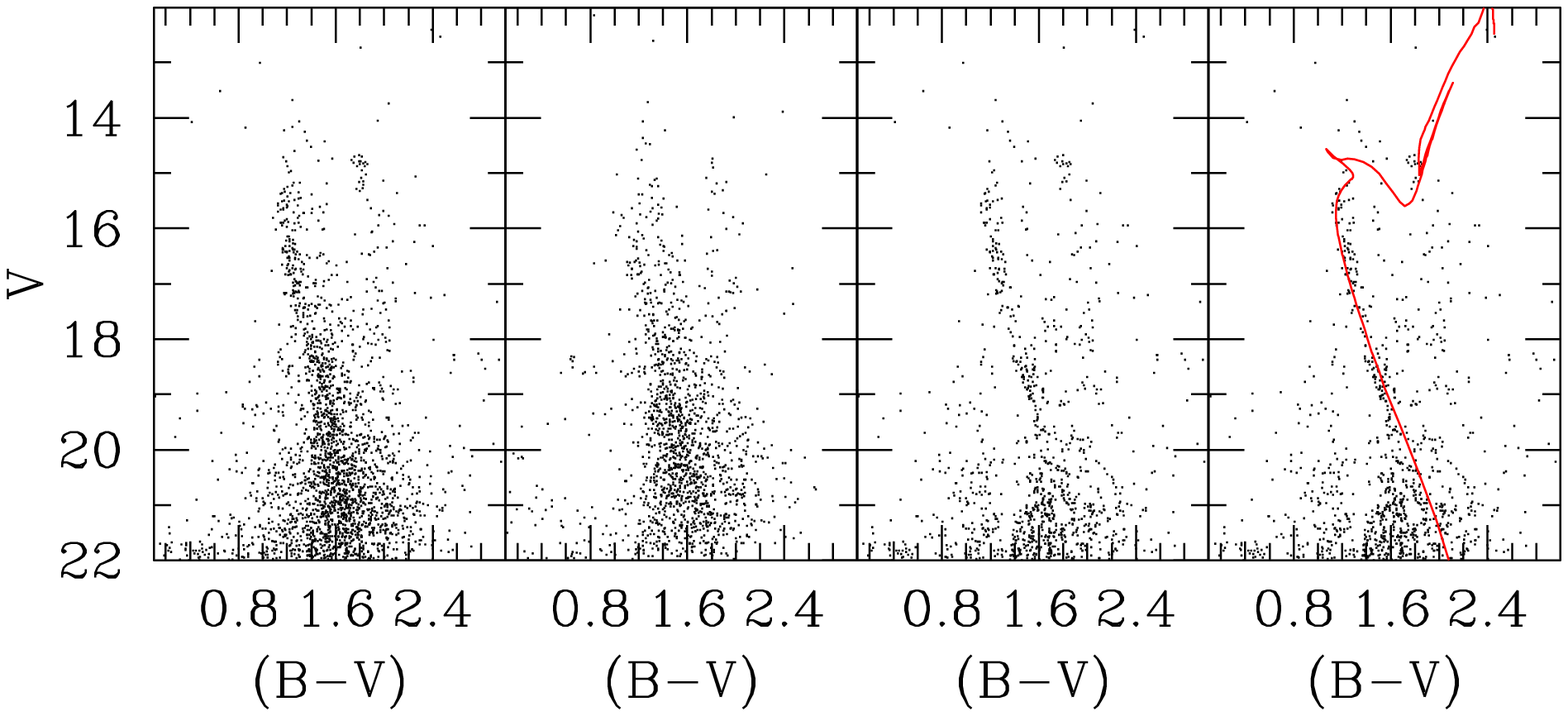} }
\caption{{\bf Left panel:} CMD of Skiff~1;{\bf Mid-left panel:}
CMD of the control field; {\bf Mid-right panel:} the clean CMD;
{\bf Right  panel:}Isochrone solution for Skiff~1: the 1.2 Gyr
isochrone is shitfed by E(B-V) = 0.85 en V-M$_V$ = 13.70}
\end{figure*} 

\begin{figure*} 
\centerline{\psfig{file=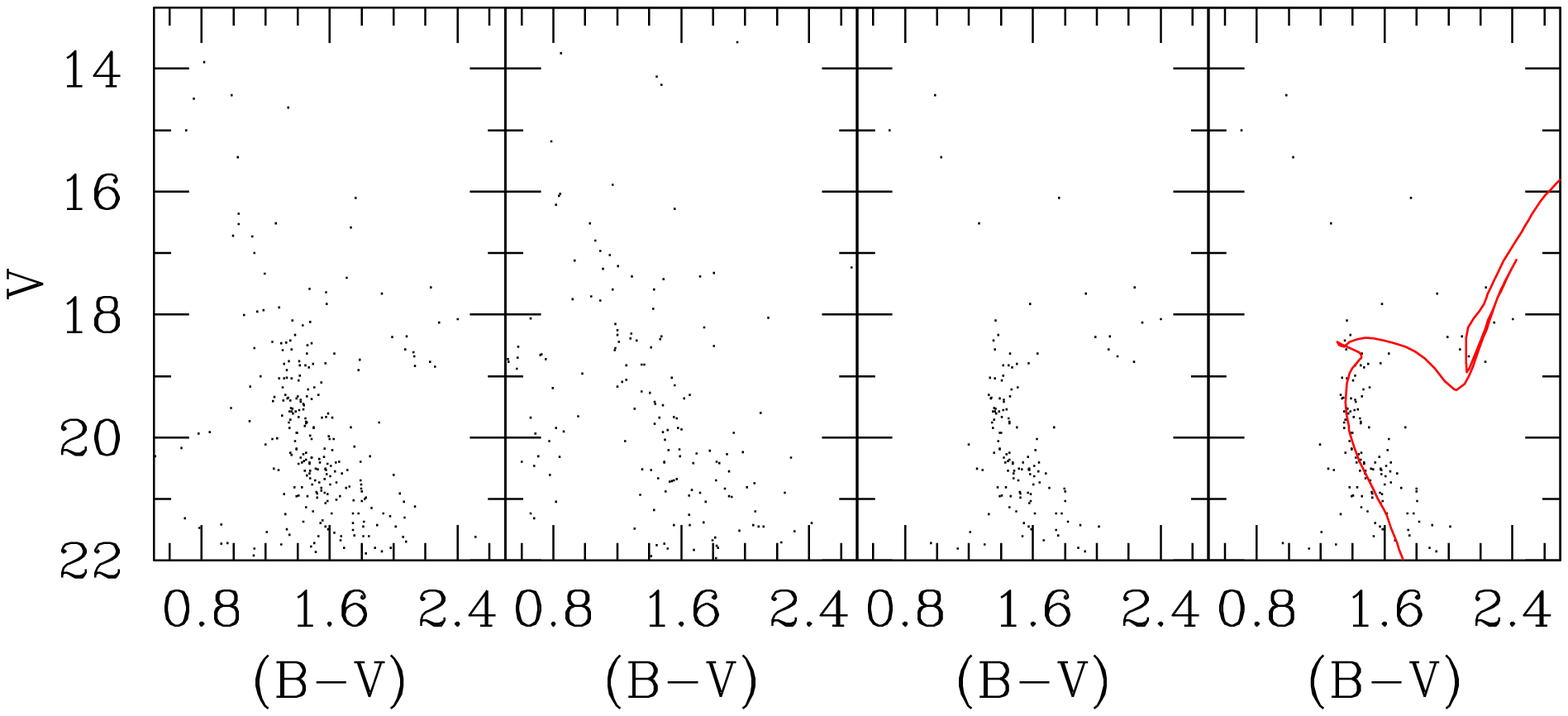} }
\caption{{\bf Left panel:} CMD of Berkeley~5;{\bf Mid-left panel:}
CMD of the control field; {\bf Mid-right panel:} the clean CMD;
{\bf Right  panel:}Isochrone solution for Berkeley~5: the 0.8 Gyr
isochrone is shitfed by E(B-V) = 1.30 en V-M$_V$ = 18.00}
\end{figure*}

\section{A closer look at Skiff 1}
We now concentrate a bit more on the open cluster Skiff~1.
It is a nearby star cluster, and although it is not a rich cluster, 
among the clusters presented here, 
the {\it cleaned} CMD is the 
most distinct, with a MS apparently extending for more than 6 mag.
This offers the opportunity to better
constrain its fundamental parameters and we employ here for this purpose
the synthetic CMD technique.\\
\noindent
The method is described in detail
in Carraro et al. (2002) and Girardi et al. (2005). 
Briefly, we count the number
of clump stars ($\sim$18) , and assign to the cluster a total mass 
($1.8 \times 10^3 M_{\odot}$) according
to the Kroupa (2001) Initial Mass Function (IMF).
A population of binaries is then added, in a 30$\%$ fraction, and with mass
ratio between 0.7 and 1.\\
Then we simulated the effect of the photometric errors, with typical values
derived from our observations.
The results are shown in Fig. 17, panels  a) and c). The age, distance
modulus, reddening and metallicity are the ones listed in Table~6.\\

\noindent
In order to estimate the location of foreground and background stars
we use a Galactic model code (Girardi et al. 2005), and generate the CMD
of the Galactic population in the direction of the cluster and within
the same cluster area. Again, this CMD is then blurred by adding
photometric errors (see panels b) and d) ).\\

\noindent
The combination of the simulated cluster and field is then shown in
panel e), which must be compared with the observations in panel f).\\

\noindent
The close similarity of the simulated and observed CMDs  ensures us
that the adopted parameters for Skiff 1 are correct within the errors,
and confirms the results of the simpler isochrone fitting method.\\

\noindent
Moreover it tells us that the Galactic model 
successfully accounts for the field population toward the cluster.
In particular the blue Main Sequence is naturally accounted for
by stars belonging to the 
of halo and thick disk of the Galaxy 
without any need to invoke an extra-population.

\begin{figure*} 
\centerline{\psfig{file=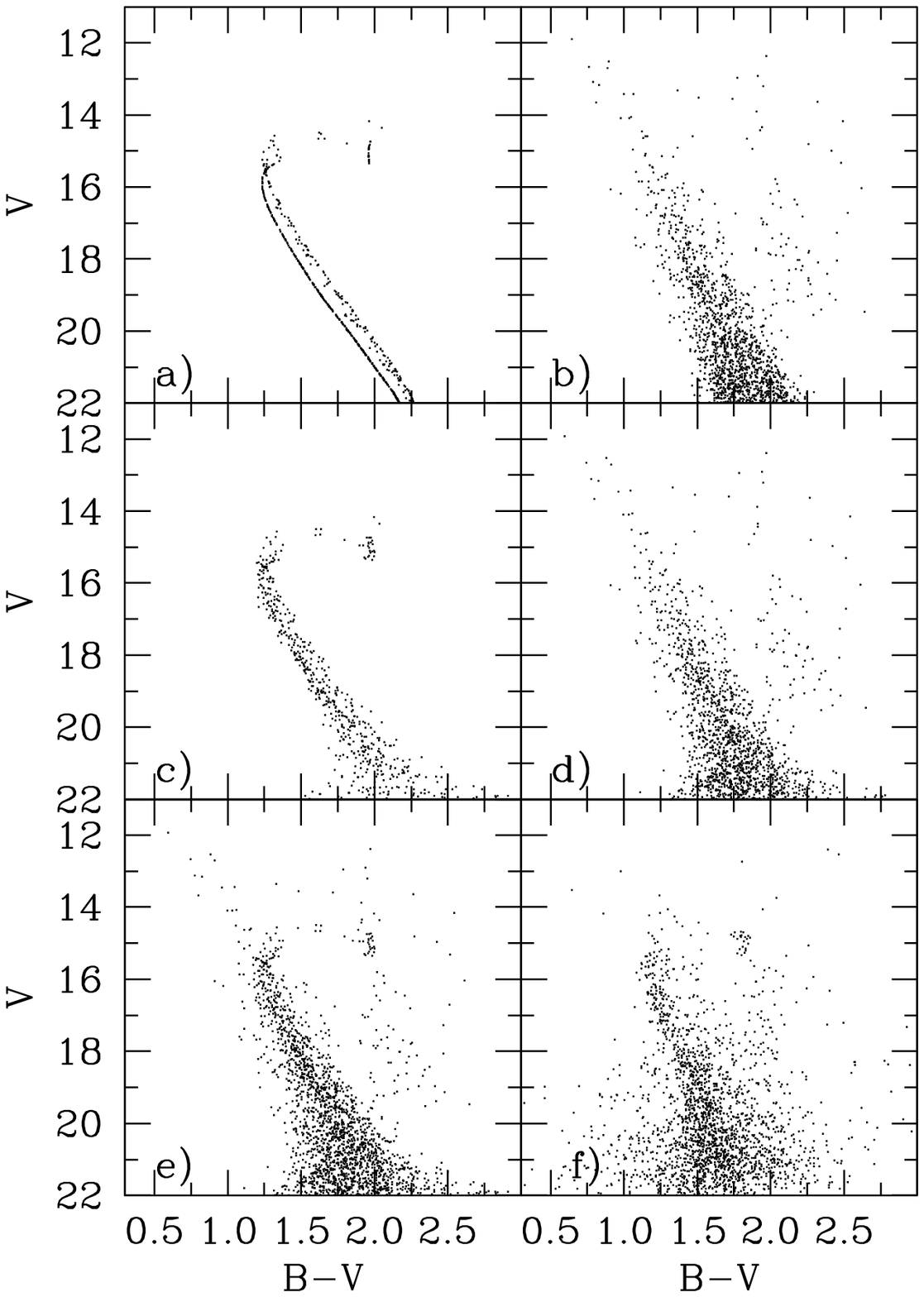}} 
\caption{Simulation of Skiff~1 and its field in the V versus B-V diagram.
{\bf (a)} Simulation of a $1.2$-Gyr old cluster with $Z=0.019$
and a initial mass of $1.8\times10^3$~$M_\odot$, based on the same
isochrones, distance modulus and colour excess as in Table~5.
We have assumed that
30 percent of the stars are binaries with mass ratios between 0.7
and 1.0.
{\bf (b)} Simulation of a $10.14 \times 10.14 {\rm arcmin}^{2}$ field centered at
Galactic coordinates $l=123^\circ.57$, $b=+5^\circ.60$, performed with
Girardi et al.\ (2005) Galactic model.
Panels {\bf (c)} and {\bf (d)} are the same as (a) and (b),
respectively, after simulation of photometric errors.
Panel {\bf (e)} shows the sum of (c) and (d), that can be
compared to the observational data shown in panel {\bf (f)}.}
\end{figure*}

\section{Discussions and Conclusions} 
We have presented CCD BVI photometry for 6 previously unstudied possibly 
old open clusters,
namely Berkeley~44, NGC~6827, Berkeley~52, Berkeley~56, Skiff~1 and Berkeley~5.\\

\noindent
We have found that all the clusters are actually old, and the ages range
from 0.8 to 4 Gyr. This sample of clusters represents an important contribution
to the poorly populated old open clusters family in the Galactic disk.
In Fig.~18 we show an updated age distribution of the old open cluster (older than 500 Myr)
so far known.\\
This comes from Carraro et al. (2005), where we added the new clusters  studied
in this paper and Auner~1 (3.5 Gyr, Carraro et al. 2006).\\

The new age distribution can be easily fitted with an exponential relation
having an e-folding time of 2 Gyr. This means than on the average the oldest
clusters in the Milky Way do not survive more than 2 Gyr.
This estimate is an order of magnitude larger than the typical life-time
of an open cluster (200 Myr), and suggests that old open clusters
survive longer possibly due to particular situations, like birth-places
high onto the Galactic plane, or the preferentially high total mass
at birth. It might also possible that some open clusters, especially in the anti-centre
could have entered the Milky Way in the past 
together with cannibalized satellites (Frinchaboy et al. 2004)\\

Much firmer conclusions might be drawn as additional old clusters are discovered
and studied.

\begin{table*}
\caption{Parameters of the studied clusters. The coordinate system
is such that
the Y axis connects the Sun to the Galactic Center, while the X axis is 
positive in the direction of galactic rotation.
Y is positive toward the Galactic anti-center, 
and X is positive in the first and 
second Galactic quadrants (Lynga 1982).}
\fontsize{8} {10pt}\selectfont
\begin{tabular}{ccccccccccc}
\hline
\multicolumn{1}{c} {$Name$} & 
\multicolumn{1}{c} {$Radius$} &
\multicolumn{1}{c} {$E(B-V)$}  &
\multicolumn{1}{c} {$(m-M)$} &
\multicolumn{1}{c} {$d_{\odot}$} &
\multicolumn{1}{c} {$X_{\odot}$} &
\multicolumn{1}{c} {$Y_{\odot}$} &
\multicolumn{1}{c} {$Z_{\odot}$} &
\multicolumn{1}{c} {$R_{GC}$} &
\multicolumn{1}{c} {$Age$} \\
\hline
& ${\prime}$& mag & mag& kpc & kpc & kpc & pc & kpc & Myr \\
\hline
Berkeley~44  & $\geq 5.0$ & 1.40$\pm$0.10  & 15.6$\pm$0.2 &  1.8 &  1.4 &  -1.1 &   100 &  7.6 & 1300$\pm$200 \\
NGC~6827     & 1.5        & 1.05$\pm$0.05  & 16.3$\pm$0.2 &  4.1 &  3.5 &  -2.1 &  -170 &  7.3 &  800$\pm$100 \\
Berkeley~52  & 1.5        & 1.50$\pm$0.10  & 18.1$\pm$0.2 &  4.9 &  4.5 &  -1.8 &  -270 &  8.1 & 2000$\pm$200 \\
Berkeley~56  & 1.0        & 0.40$\pm$0.05  & 16.6$\pm$0.2 & 12.1 & 12.0 &  -0.8 & -1100 & 14.3 & 4000$\pm$400 \\
Skiff~1      & $\geq$5.0  & 0.85$\pm$0.05  & 13.7$\pm$0.2 &  1.6 &  1.3 &   0.9 &   160 &  9.5 & 1200$\pm$100 \\
Berkeley~5   & 1.0        & 1.30$\pm$0.10  & 18.0$\pm$0.2 &  6.2 &  4.8 &   3.9 &    80 & 13.3 &  800$\pm$100 \\
\hline
\end{tabular}
\end{table*}

\section*{Acknowledgements} 
The work of G. Carraro is supported by {\it Fundaci\'on Andes}.
This study made use of Simbad and WEBDA databases.

\begin{figure} 
\centerline{\psfig{file=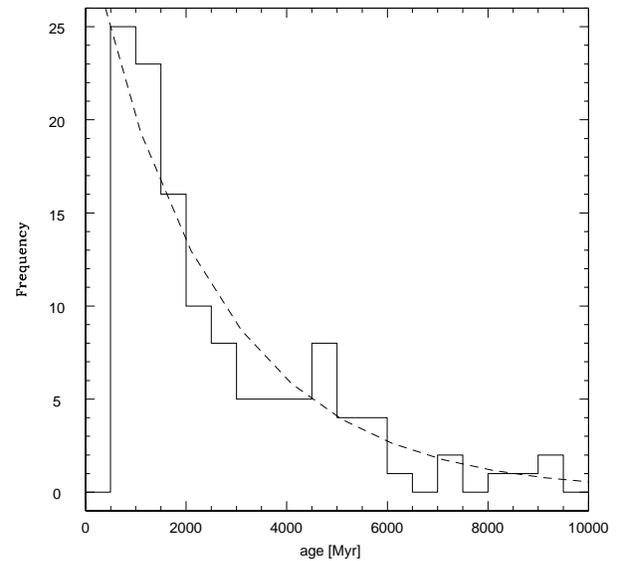,width=\columnwidth}} 
\caption{Age distribution of known open clusters older than 500 Myr.
The dashed exponential line has an e-folding time of 2 Gyr.}
\end{figure}

\end{document}